\documentclass[floatfix,showpacs]{revtex4}
\usepackage{graphicx}
\usepackage{bm}

\begin{document}

\title{Collective oscillations of a confined Bose gas at finite
temperature in the random-phase approximation}
\author{Xia-Ji Liu}
\affiliation{\ Department of Physics, Tsinghua University, Beijing
100084, China}
\author{Hui Hu}
\affiliation{\ NEST-INFM and Classe di Scienze, \\
Scuola Normale Superiore, I-56126 Pisa, Italy}
\author{A. Minguzzi}
\affiliation{\ NEST-INFM and Classe di Scienze, \\
Scuola Normale Superiore, I-56126 Pisa, Italy}
\author{M. P. Tosi}
\affiliation{\ NEST-INFM and Classe di Scienze, \\
Scuola Normale Superiore, I-56126 Pisa, Italy}
\date{\today}

\begin{abstract}
We present a theory for the linear dynamics of a weakly
interacting Bose gas confined inside a harmonic trap at finite
temperature. The theory treats the motions of the condensate and
of the non-condensate on an equal footing within a generalized
random-phase approximation, which ({\it i}) extends the
second-order Beliaev-Popov approach by allowing for the dynamical
coupling between fluctuations in the thermal cloud, and ({\it ii})
reduces to an earlier random-phase scheme when the anomalous
density fluctuations are omitted. Numerical calculations of the
low-lying spectra in the case of isotropic confinement show that
the present theory obeys with high accuracy the generalized Kohn
theorem for the dipolar excitations and demonstrate that combined
normal and anomalous density fluctuations play an important role
in the monopolar excitations of the condensate. Mean-field theory
is instead found to yield accurate results for the quadrupolar
modes of the condensate. Although the restriction to spherical
confinement prevents quantitative comparisons with measured
spectra, it appears that the non-mean field effects that we
examine may be relevant to explain the features exhibited by the
breathing mode as a function of temperature in the experiments
carried out at JILA on a gas of $^{87}$Rb atoms.
\end{abstract}

\maketitle

\section{Introduction}
Soon after the realization of Bose-Einstein condensation in
trapped atomic gases, an important development in this field has
been the measurement of the frequencies and damping rates of
collective excitations \cite
{jin96,jin97,ketterle96,ketterle98,ketterle00}. These measurements
are very accurate and provide a unique opportunity for
quantitative tests of quantum theories of the dynamics of
many-body systems. In particular, the measurements of the
lowest-energy excitations made at JILA \cite{jin97} on $^{87}$Rb
gases at various temperatures have proved hard to understand at
simple mean-field level \cite{burnett,hutchinson} and have
therefore stimulated a number of theoretical studies to address
effects beyond the mean-field approximation \cite
{overview,fs,giorgini,morgan00,morgan03,reidl00,stoof,zgn,mt}.

The key issue in investigations transcending the mean-field level
is the {\em full} dynamic description of both condensed and
non-condensed atoms and their mutual interactions \cite{mt}. While
the condensate dynamics is well described by a single nonlinear
Gross-Pitaevskii equation (GPE), how to monitor the evolution of
the non-condensate is a much more delicate problem. The best
candidate theory that takes into account the coupled dynamics of
condensate and non-condensate for a homogeneous weakly-interacting
Bose gas in the collisionless limit is the second-order
Beliaev-Popov (SOBP) theory \cite {beliaev}, which has been
reexamined recently by Shi and Griffin \cite {griffin98} and
extended to trapped gases by Fedichev and Shlyapnikov \cite{fs}
and by Giorgini \cite{giorgini} (see also Rusch {\it et al.}
\cite{morgan00}). However, for the trapped gas the Thomas-Fermi
approximation on the SOBP theory fails to account for the JILA
observations \cite{fs,giorgini}. One possible reason is that the
dynamics of the condensate and non-condensate are not treated on
an equal footing in the theory, {\it i.e.} the dynamical coupling
between fluctuations in the thermal cloud is not included. This
coupling should be important when the thermal fraction is
significantly populated and, as will be discussed below, is in
fact needed to satisfy the generalized Kohn theorem for the dipole
modes. One way to include these processes is to use the linear
response theory in the random-phase approximation (RPA) as
developed by two of us \cite{mt}. Such a treatment chooses the
Hartree-Fock gas as the reference system for the thermal atoms,
thus neglecting the anomalous density fluctuations that may play a
role at intermediate temperatures.

In the present paper we improve on the Hartree-Fock RPA (HF-RPA)
by including the anomalous density fluctuations. The resulting
theory can be referred to as the HFB-RPA since our choice of the
reference system is provided by the first-order
Hartree-Fock-Bogoliubov theory. We explicitly show that the
HFB-RPA theory formally reduces to the SOBP theory given by
Fedichev and Shlyapnikov \cite{fs} and by Giorgini \cite{giorgini}
if (i) one excludes the process of driving the non-condensate by
its self-generated dynamical potential, and (ii) one keeps only
terms of second order in the coupling constant. It is interesting
to note that the HF-RPA similarly reduces to the dielectric
formalism given by Reidl {\it et al}. \cite{reidl00}.

We then numerically investigate the low-lying excitations of a
fluid representing a Bose-condensed gas of 2000 $^{87}$Rb atoms in
a spherically symmetric harmonic trap at finite temperature by
using the HFB-RPA as well as the SOBP theory and the HF-RPA. All
three theories give qualitatively the same results for the
quadrupolar mode of the condensate. However they predict different
trends for the monopolar mode, due to the strong coupling between
the oscillations of the condensate and those of the
non-condensate. For the first time this observation highlights the
crucial roles played already in the linear excitation spectra by
the normal and anomalous density fluctuations of the
non-condensate.

The paper is organized as follows. In Sec. II we derive the
generalized RPA equations within the framework of the
Hartree-Fock-Bogoliubov approximation, and in Sec. III we briefly
demonstrate how to deduce the SOBP theory from the HFB-RPA
equations. In Sec. IV we describe our numerical procedure for
calculating the spectral response functions and check their
accuracy, and in Secs. V and VI we present our numerical results
for the low-energy excitations. Finally, Sec. VII presents our
main conclusions.

\section{The HFB-RPA theory}

The essential idea of the RPA is that the gas responds as a
reference gas to self-consistent dynamical potentials
\cite{overview}. In the HF-RPA
treatment one chooses as dynamical variables the density fluctuations $%
\delta n_c$ of the condensate and $\delta \tilde{n}$ of the
non-condensate \cite{mt}. The HF-RPA equations follow by imposing
that the condensed and non-condensed particles experience
dynamical Hartree-Fock potentials generated by both types of
density fluctuations and respond to them as a Hartree-Fock gas.

Our starting point for the derivation of the HFB-RPA is the
definition of the appropriate single-particle reference system.
The contribution of the anomalous density is included by choosing
the Hartree-Fock-Bogoliubov gas at finite temperature as
reference, which is defined in terms of the condensate
wavefunction $\Phi _0$ and of the single-particle amplitudes $u_j$
and $v_j$ for the non-condensate \cite{griffin96}. The condensate
is described by the generalized GPE
\begin{equation}
\left[ -\frac{\hbar ^2\mathbf{\bigtriangledown
}^2}{2m}+V_{ext}\left( \mathbf{r}\right) +g\left( n_c\left(
\mathbf{r}\right) +2\tilde{n}^0\left(
\mathbf{r}\right) \right) \right] \Phi _0\left( \mathbf{r}\right) +g\tilde{m}%
^0\left( \mathbf{r}\right) \Phi _0^{*}\left( \mathbf{r}\right)
=\mu \Phi _0\left( \mathbf{r}\right) ,  \label{GPE}
\end{equation}
where we adopt the standard contact-pseudopotential model
characterized by
the coupling constant $g=4\pi \hbar ^2a/m$, with $a$ being the \textit{s}%
-wave scattering length. In Eq. (\ref{GPE}) $V_{ext}\left(
\mathbf{r}\right) =m(\omega _x^2x^2+\omega _y^2y^2+\omega
_z^2z^2)/2$ is the external
confinement and $n_c\left( \mathbf{r}\right) =\left| \Phi _0\left( \mathbf{r}%
\right) \right| ^2$, $\tilde{n}^0\left( \mathbf{r}\right)
=\sum_j\left[ \left( \left| u_j\left( \mathbf{r}\right) \right|
^2+\left| v_j\left( \mathbf{r}\right) \right| ^2\right) f_j+\left|
v_j\left( \mathbf{r}\right)
\right| ^2\right] $, and $\tilde{m}^0\left( \mathbf{r}\right) =$ $%
\sum_j\left[ \left( 1+2f_j\right) u_j\left( \mathbf{r}\right)
v_j^{*}\left( \mathbf{r}\right) \right] $ are the condensate
density and the normal and anomalous thermal densities,
$f_j=1/\left( e^{\beta \epsilon _j}-1\right) $ being the
Bose-Einstein distribution with $\beta =1/k_BT$ and $\mu $ the
chemical potential. The non-condensate amplitudes are obtained by
the solution of the generalized Bogoliubov-deGennes equations
\begin{equation}
\left\{
\begin{array}{l}
\mathcal{L}\left( \mathbf{r}\right) u_j\left( \mathbf{r}\right)
+g\left( \Phi _0^2\left( \mathbf{r}\right) +\tilde{m}^0\left(
\mathbf{r}\right) \right) v_j\left( \mathbf{r}\right) =\epsilon
_ju_j\left( \mathbf{r}\right)
\\
\mathcal{L}\left( \mathbf{r}\right) v_j\left( \mathbf{r}\right)
+g\left( \Phi _0^{*2}\left( \mathbf{r}\right)
+\tilde{m}^{0*}\left( \mathbf{r}\right) \right) u_j\left(
\mathbf{r}\right) =-\epsilon _jv_j\left( \mathbf{r}\right) .
\end{array}
\right.   \label{BdG}
\end{equation}
Here $\mathcal{L}\left( \mathbf{r}\right) =-\hbar
^2\mathbf{\bigtriangledown }^2/2m+V_{ext}\left( \mathbf{r}\right)
+2g\left( n_c\left( \mathbf{r}\right) +\tilde{n}^0\left(
\mathbf{r}\right) \right) $. The Popov approximation to
the Hartree-Fock-Bogoliubov (HFB-Popov) theory is recovered by setting $%
\tilde{m}^0\left( \mathbf{r}\right) =0$ in Eqs. (\ref{GPE}) and
(\ref{BdG}) \cite{griffin96}.

We would like to remark that from a dynamical point of view the amplitudes $%
u_j$ and $v_j$ can alternatively be viewed as excitations out of
the condensate. The duality of such mean-field description follows
from the assumption of Bose symmetry breaking (see \textit{e.g.}
\cite{rmp} for a discussion).

In deriving next the HFB-RPA equations we adopt five dynamic
variables, which are the fluctuations $\delta \Phi $ and $\delta
\Phi ^{*}$ of the condensate wavefunction and its complex
conjugate, the normal density
fluctuation $\delta \tilde{n}$, and the anomalous density fluctuation $%
\delta \tilde{m}$ together with its complex conjugate $\delta \tilde{m}^{*}$%
. $\delta \Phi $ and $\delta \Phi ^{*}$ are separately introduced
because of their different coupling to $\delta \tilde{m}$ and
$\delta \tilde{m}^{*}$, and are related to the density fluctuation
of the condensate by $\delta n_c=\Phi _0^{*}\delta \Phi +\Phi
_0\delta \Phi ^{*}$. The HFB-RPA then follows naturally by
evaluating the self-consistent dynamical Hartree-Fock-Bogoliubov
potential generated by the density fluctuations of the condensate
(phonon quasiparticles) and of the non-condensate (thermal
quasiparticles). This can be done by invoking the decomposition
\begin{equation}
\psi \left( \mathbf{r},t\right) =\Phi _0\left( \mathbf{r}\right) +\tilde{\psi%
}\left( \mathbf{r},t\right)   \label{BSB}
\end{equation}
for the Bose field operator in the interaction Hamiltonian,
\begin{eqnarray}
\mathcal{H}_{int} &=&\frac g2\int d\mathbf{r}\psi ^{+}\left( \mathbf{r}%
,t\right) \psi ^{+}\left( \mathbf{r},t\right) \psi \left( \mathbf{r}%
,t\right) \psi \left( \mathbf{r},t\right)   \nonumber \\
&=&\frac g2\int d\mathbf{r}\left[ \left| \Phi _0\right| ^4+2\left|
\Phi
_0\right| ^2\Phi _0^{*}\tilde{\psi}+2\left| \Phi _0\right| ^2\Phi _0\tilde{%
\psi}^{+}\right.   \nonumber \\
&&\left. +\Phi _0^{*}\Phi _0^{*}\tilde{\psi}\tilde{\psi}+4\left|
\Phi
_0\right| ^2\tilde{\psi}^{+}\tilde{\psi}+\Phi _0\Phi _0\tilde{\psi}^{+}%
\tilde{\psi}^{+}\right.   \nonumber \\
&&\left. +2\Phi _0^{*}\tilde{\psi}^{+}\tilde{\psi}\tilde{\psi}+2\Phi _0%
\tilde{\psi}^{+}\tilde{\psi}^{+}\tilde{\psi}+\tilde{\psi}^{+}\tilde{\psi}^{+}%
\tilde{\psi}\tilde{\psi}\right] \text{.}  \label{intHami}
\end{eqnarray}
Note that in the choice made in Eq. (\ref{BSB}), which is
different from those generally used in the literature, the
\emph{non-equilibrium} statistical average $\left\langle
\tilde{\psi}\left( \mathbf{r},t\right) \right\rangle $ of the
operator $\tilde{\psi}\left( \mathbf{r},t\right) $ is non-zero
since we prefer to extract from $\psi \left( \mathbf{r},t\right) $
a \emph{time-independent} condensate wavefunction. Rather, $\tilde{\psi}%
\left( \mathbf{r},t\right) $ gives the field operator for the
phonon quasiparticles and describes the condensate fluctuation,
$\left\langle \tilde{\psi}\left( \mathbf{r},t\right) \right\rangle
=\left\langle \psi \left( \mathbf{r},t\right) \right\rangle -\Phi
_0\left( \mathbf{r}\right) =\Phi \left( \mathbf{r},t\right) -\Phi
_0\left( \mathbf{r}\right) =\delta \Phi \left( \mathbf{r},t\right)
.$ Analogously $\delta \Phi ^{*}\left( \mathbf{r},t\right)
=\left\langle \psi ^{+}\left( \mathbf{r},t\right) \right\rangle
-\Phi _0^{*}\left( \mathbf{r}\right) $.

The self-consistent dynamical potentials are originated from the \emph{%
higher-order} correlation terms beyond the mean-field description
and are contained in the last line of Eq. (\ref{intHami}). We
approximate these terms by using Wick's theorem in the following
manner:
\begin{eqnarray}
2\Phi _0^{*}\tilde{\psi}^{+}\tilde{\psi}\tilde{\psi} &\simeq
&4\Phi
_0^{*}\left\langle \tilde{\psi}^{+}\tilde{\psi}\right\rangle \tilde{\psi}%
+2\Phi _0^{*}\left\langle \tilde{\psi}\tilde{\psi}\right\rangle \tilde{\psi}%
^{+}+4\Phi _0^{*}\delta \Phi \tilde{\psi}^{+}\tilde{\psi}+2\Phi
_0^{*}\delta
\Phi ^{*}\tilde{\psi}\tilde{\psi},  \label{D1} \\
2\Phi _0\tilde{\psi}^{+}\tilde{\psi}^{+}\tilde{\psi} &\simeq
&4\Phi
_0\left\langle \tilde{\psi}^{+}\tilde{\psi}\right\rangle \tilde{\psi}%
^{+}+2\Phi _0\left\langle
\tilde{\psi}^{+}\tilde{\psi}^{+}\right\rangle \tilde{\psi}+4\Phi
_0\delta \Phi ^{*}\tilde{\psi}^{+}\tilde{\psi}+2\Phi _0\delta \Phi
\tilde{\psi}^{+}\tilde{\psi}^{+}  \label{D2}
\end{eqnarray}
and
\begin{equation}
\tilde{\psi}^{+}\tilde{\psi}^{+}\tilde{\psi}\tilde{\psi}\simeq
4\left\langle
\tilde{\psi}^{+}\tilde{\psi}\right\rangle \tilde{\psi}^{+}\tilde{\psi}%
+\left\langle \tilde{\psi}^{+}\tilde{\psi}^{+}\right\rangle \tilde{\psi}%
\tilde{\psi}+\left\langle \tilde{\psi}\tilde{\psi}\right\rangle \tilde{\psi}%
^{+}\tilde{\psi}^{+}.  \label{D3}
\end{equation}
Explicitly, the fluctuations of the non-condensate are defined by
$\delta
\tilde{n}\left( \mathbf{r},t\right) =\left\langle \psi ^{+}\left( \mathbf{r}%
,t\right) \psi \left( \mathbf{r},t\right) \right\rangle
-\tilde{n}^0\left( \mathbf{r}\right) $, $\delta \tilde{m}\left(
\mathbf{r},t\right)
=\left\langle \psi \left( \mathbf{r},t\right) \psi \left( \mathbf{r}%
,t\right) \right\rangle -\tilde{m}^0\left( \mathbf{r}\right) $,
and $\delta \tilde{m}^{*}\left( \mathbf{r},t\right) =\left\langle
\psi ^{+}\left(
\mathbf{r},t\right) \psi ^{+}\left( \mathbf{r},t\right) \right\rangle -%
\tilde{m}^{0*}\left( \mathbf{r}\right) $. We insert these
definitions into
Eqs. (\ref{D1})-(\ref{D3}), remove the terms that are proportional to $%
\tilde{n}^0\left( \mathbf{r}\right) $, $\tilde{m}^0\left(
\mathbf{r}\right) $ and $\tilde{m}^{0*}\left( \mathbf{r}\right) $
as these are already accounted by the Hartree-Fock-Bogoliubov
mean-field equations, and finally collect together the remaining
terms. We then find that the self-consistent dynamical potential
induced by fluctuations is
\begin{eqnarray}
\mathcal{\delta V}^{SC} &=&g\int d\mathbf{r}\left[ 2\Phi
_0^{*}\delta \tilde{n}\tilde{\psi}+\Phi _0^{*}\delta
\tilde{m}\tilde{\psi}^{+}+2\Phi
_0\delta \tilde{n}\tilde{\psi}^{+}+\Phi _0\delta \tilde{m}^{*}\tilde{\psi}%
\right.   \nonumber \\
&&\left. +2\Phi _0^{*}\delta \Phi
\tilde{\psi}^{+}\tilde{\psi}+\Phi
_0^{*}\delta \Phi ^{*}\tilde{\psi}\tilde{\psi}+2\Phi _0\delta \Phi ^{*}%
\tilde{\psi}^{+}\tilde{\psi}+\Phi _0\delta \Phi \tilde{\psi}^{+}\tilde{\psi}%
^{+}\right.   \nonumber \\
&&\left. +2\delta \tilde{n}\tilde{\psi}^{+}\tilde{\psi}+\delta \tilde{m}%
\tilde{\psi}^{+}\tilde{\psi}^{+}/2+\delta \tilde{m}^{*}\tilde{\psi}\tilde{%
\psi}/2\right] \text{.}  \label{scHFB}
\end{eqnarray}
Physically, the eight leading terms on the right-hand side of Eq.
(\ref {scHFB}) are the self-consistent potentials generated by
phonon quasiparticles on themselves and on thermal quasiparticles.
These terms have been discussed by Giorgini \cite{giorgini} and by
Liu and Hu \cite{liu} and, as we shall see explicitly below, lead
to the SOBP theory in a perturbative treatment to second order in
the coupling constant. On the other hand, the last three terms in
Eq. (\ref{scHFB}) describe the self-potential of the thermal
quasiparticles and are expected to excite zero-sound-like
collective modes of the non-condensate. Although these terms are
only of third order in the coupling constant and therefore are
missing in the SOBP theory, they may have a significant role when
the depletion of the condensate is large. They are also required
for consistency with the generalized Kohn theorem.

With the self-consistent Hartree-Fock-Bogoliubov potential in Eq.
(\ref {scHFB}) and using the notation $\chi f\equiv \int
d\mathbf{r}^{\prime }\chi \left( \mathbf{r},\mathbf{r}^{\prime
},\omega \right) f(\mathbf{r}^{\prime }) $, we can write the
coupled HFB-RPA equations for $\delta \Phi $, $\delta \Phi ^{*}$,
$\delta \tilde{n}$, $\delta \tilde{m}$ and $\delta \tilde{m}^{*}$
in a compact matrix form. Within the linear response framework we
have
\begin{equation}
\left(
\begin{array}{c}
\delta \Phi  \\
\delta \Phi ^{*}
\end{array}
\right) =g\left(
\begin{array}{cc}
\chi _{cc} & \chi _{c\bar{c}} \\
\chi _{\bar{c}c} & \chi _{\bar{c}\bar{c}}
\end{array}
\right) \left(
\begin{array}{c}
2\Phi _0^{*}\delta \tilde{n}+\Phi _0\delta \tilde{m}^{*} \\
2\Phi _0\delta \tilde{n}+\Phi _0^{*}\delta \tilde{m}
\end{array}
\right)   \label{HFBRPA1}
\end{equation}
and
\begin{equation}
\left(
\begin{array}{c}
\delta \tilde{n} \\
\delta \tilde{m} \\
\delta \tilde{m}^{*}
\end{array}
\right) =g\left(
\begin{array}{ccc}
\chi _{\tilde{n}\tilde{n}} & \chi _{\tilde{n}\tilde{m}} & \chi _{\tilde{n}%
\tilde{m}^{+}} \\
\chi _{\tilde{m}\tilde{n}} & \chi _{\tilde{m}\tilde{m}} & \chi _{\tilde{m}%
\tilde{m}^{+}} \\
\chi _{\tilde{m}^{+}\tilde{n}} & \chi _{\tilde{m}^{+}\tilde{m}} & \chi _{%
\tilde{m}^{+}\tilde{m}^{+}}
\end{array}
\right) \left(
\begin{array}{c}
2\Phi _0^{*}\delta \Phi +2\Phi _0\delta \Phi ^{*}+2\delta \tilde{n} \\
\Phi _0^{*}\delta \Phi ^{*}+\delta \tilde{m}^{*}/2 \\
\Phi _0\delta \Phi +\delta \tilde{m}/2
\end{array}
\right) .  \label{HFBRPA2}
\end{equation}
In these equations $\chi _{\alpha \beta }$ ($\alpha $, $\beta =c$
or $\bar{c} $ ) and $\chi _{ab}$ ($a$, $b=\tilde{n}$, $\tilde{m}$,
or $\tilde{m}^{+}$) are the two-particle response functions of the
condensate and non-condensate components, respectively. They can
easily be evaluated by using the quasiparticle amplitudes obtained
from the Hartree-Fock-Bogoliubov solutions
in the standard finite-temperature Green's functions technique \cite{fetter}%
. For the condensate we have
\begin{eqnarray}
\chi _{cc}\left( \mathbf{r},\mathbf{r}^{\prime },\omega \right)
&=&\sum_j\left( \frac{u_j\left( \mathbf{r}\right) v_j^{*}\left( \mathbf{r}%
^{\prime }\right) }{\hbar \omega ^{+}-\epsilon
_j}-\frac{v_j^{*}\left( \mathbf{r}\right) u_j\left(
\mathbf{r}^{\prime }\right) }{\hbar \omega
^{+}+\epsilon _j}\right) ,  \label{kapacccc} \\
\chi _{c\bar{c}}\left( \mathbf{r},\mathbf{r}^{\prime },\omega
\right)
&=&\sum_j\left( \frac{u_j\left( \mathbf{r}\right) u_j^{*}\left( \mathbf{r}%
^{\prime }\right) }{\hbar \omega ^{+}-\epsilon
_j}-\frac{v_j^{*}\left( \mathbf{r}\right) v_j\left(
\mathbf{r}^{\prime }\right) }{\hbar \omega
^{+}+\epsilon _j}\right) ,  \label{kapaccct} \\
\chi _{\bar{c}c}\left( \mathbf{r},\mathbf{r}^{\prime },\omega
\right)
&=&\sum_j\left( \frac{v_j\left( \mathbf{r}\right) v_j^{*}\left( \mathbf{r}%
^{\prime }\right) }{\hbar \omega ^{+}-\epsilon
_j}-\frac{u_j^{*}\left( \mathbf{r}\right) u_j\left(
\mathbf{r}^{\prime }\right) }{\hbar \omega ^{+}+\epsilon
_j}\right)   \label{kapactcc}
\end{eqnarray}
and
\begin{equation}
\chi _{\bar{c}\bar{c}}\left( \mathbf{r},\mathbf{r}^{\prime
},\omega \right)
=\sum_j\left( \frac{v_j\left( \mathbf{r}\right) u_j^{*}\left( \mathbf{r}%
^{\prime }\right) }{\hbar \omega ^{+}-\epsilon
_j}-\frac{u_j^{*}\left( \mathbf{r}\right) v_j\left(
\mathbf{r}^{\prime }\right) }{\hbar \omega ^{+}+\epsilon
_j}\right) ,  \label{kapactct}
\end{equation}
where $\omega ^{+}=\omega +i\eta $ with $\eta =0^{+}$. The
expressions for the two-particle response functions of the
non-condensate are lengthier and we list them in Appendix A.

The coupled HFB-RPA equations (\ref{HFBRPA1}) and (\ref{HFBRPA2})
are the central result of this work. They reduce to the HF-RPA
equations if one omits the anomalous density fluctuations of
thermal quasiparticles. That is, the HF-RPA gives
\begin{equation}
\delta n_c\left( \mathbf{r},\omega \right) =2g\int
d\mathbf{r}^{\prime }\chi
_c\left( \mathbf{r},\mathbf{r}^{\prime };\omega \right) \delta \tilde{n}(%
\mathbf{r}^{\prime },\omega )  \label{HFRPA1}
\end{equation}
and
\begin{equation}
\delta \tilde{n}\left( \mathbf{r},\omega \right) =2g\int
d\mathbf{r}^{\prime }\chi _{\tilde{n}\tilde{n}}\left(
\mathbf{r},\mathbf{r}^{\prime };\omega \right) \left[ \delta
n_c\left( \mathbf{r}^{\prime },\omega \right) +\delta
\tilde{n}(\mathbf{r}^{\prime },\omega )\right] ,  \label{HFRPA2}
\end{equation}
where $\chi _c\left( \mathbf{r},\mathbf{r}^{\prime };\omega
\right) =\Phi _0^{*}\left( \mathbf{r}\right) \chi _{cc}\Phi
_0^{*}(\mathbf{r}^{\prime
})+\Phi _0\left( \mathbf{r}\right) \chi _{\bar{c}c}\Phi _0^{*}(\mathbf{r}%
^{\prime })+\Phi _0^{*}\left( \mathbf{r}\right) \chi _{c\bar{c}}\Phi _0(%
\mathbf{r}^{\prime })+\Phi _0\left( \mathbf{r}\right) \chi _{\bar{c}\bar{c}%
}\Phi _0(\mathbf{r}^{\prime })$. One must accordingly take the
Hartree-Fock reference system in the calculation of the response
functions \cite{mt}.

\section{Reduction to the second-order Beliaev-Popov theory}

In this Section we show that the coupled HFB-RPA equations for the
normal modes of the condensate simplify to those obtained in the
SOBP theory if we neglect the self-coupling of density
fluctuations in the non-condensate and keep only terms up to
second order in the coupling constant $g$. This discussion also
allows us to define a RPA form of the SOBP theory, that will later
be used in our numerical calculations.

If we neglect the terms in $\delta \tilde{n}$, $\delta \tilde{m}$ and $%
\delta \tilde{m}^{*}$ on the right-hand side of Eq.
(\ref{HFBRPA2}) and substitute this equation in Eq.
(\ref{HFBRPA1}), we immediately obtain the self-consistent
equations for the fluctuations of the condensate as
\begin{equation}
\left(
\begin{array}{l}
\delta \Phi  \\
\delta \Phi ^{*}
\end{array}
\right) =g^2\left(
\begin{array}{cc}
\chi _{cc} & \chi _{c\bar{c}} \\
\chi _{\bar{c}c} & \chi _{\bar{c}\bar{c}}
\end{array}
\right) \mathcal{D}\left(
\begin{array}{l}
\delta \Phi  \\
\delta \Phi ^{*}
\end{array}
\right) ,  \label{sobp1}
\end{equation}
where the matrix $\mathcal{D}$ is defined as
\begin{equation}
\mathcal{D}=\left(
\begin{array}{lll}
2\Phi _0^{*} & 0 & \Phi _0 \\
2\Phi _0 & \Phi _0^{*} & 0
\end{array}
\right) \left(
\begin{array}{lll}
\chi _{\tilde{n}\tilde{n}} & \chi _{\tilde{n}\tilde{m}} & \chi _{\tilde{n}%
\tilde{m}^{+}} \\
\chi _{\tilde{m}\tilde{n}} & \chi _{\tilde{m}\tilde{m}} & \chi _{\tilde{m}%
\tilde{m}^{+}} \\
\chi _{\tilde{m}^{+}\tilde{n}} & \chi _{\tilde{m}^{+}\tilde{m}} & \chi _{%
\tilde{m}^{+}\tilde{m}^{+}}
\end{array}
\right) \left(
\begin{array}{ll}
2\Phi _0^{*} & 2\Phi _0 \\
0 & \Phi _0^{*} \\
\Phi _0 & 0
\end{array}
\right) .  \label{matrixd}
\end{equation}
Equation (\ref{sobp1}) is already of second order in $g$ and we
shall regard it as providing a second-order Beliaev-Popov theory
within a random-phase framework (SOBP-RPA).

The SOBP-RPA differs only slightly from the SOBP theory presented
in Ref. \cite{giorgini}, in the sense that it still keeps a class
of terms beyond second order. In fact, to second order in the
coupling constant we can describe the small oscillations of the
condensate by a set $(u_{osc},v_{osc}) $ of quasiparticle
amplitudes with excitation energy $\epsilon _{osc}$. By setting
$(\delta \Phi ,\delta \Phi ^{*})=(u_{osc},v_{osc})$ in Eq. (\ref
{sobp1}) and using Eqs. (\ref{kapacccc})-(\ref{kapactct}) we find
the eigenfrequency of the oscillations of the condensate as
$\epsilon _{osc}+\delta E-i\gamma ,$ where
\begin{eqnarray}
\delta E-i\gamma &=&g^2\left(
\begin{array}{ll}
v_{osc}^{*}, & u_{osc}^{*}
\end{array}
\right) \mathcal{D}\left(
\begin{array}{l}
u_{osc} \\
v_{osc}
\end{array}
\right)  \nonumber \\
&=&g\int d\mathbf{r}\Phi _0\left[ 2\left(
u_{osc}^{*}+v_{osc}^{*}\right)
\delta \tilde{n}+u_{osc}^{*}\delta \tilde{m}+v_{osc}^{*}\delta \tilde{m}%
^{*}\right] .  \label{deltae}
\end{eqnarray}
In recent work two of us \cite{liu} have explicitly shown that Eq.
(\ref {deltae}) agrees with the result for the eigenfrequency
shift given by the SOBP theory of Giorgini \cite{giorgini}.

\section{Numerical procedure}

We turn to numerical illustrations of the excitation spectra with
the main aim of comparatively examining the three theoretical
approaches that we have introduced in Secs. II and III. We do this
in the case of a spherically symmetric trap in view of the
complexity of the calculations involved.

We excite density fluctuations by applying a time-dependent
perturbation of the form $F(t)\propto \exp \left( i\omega t\right)
\int d\mathbf{r}V_p\left(
\mathbf{r}\right) \psi ^{+}\left( \mathbf{r}\right) \psi \left( \mathbf{r}%
\right) $. In the HFB-RPA this corresponds to adding the terms
$\left[
\begin{array}{ll}
\chi _{cc}V_p\Phi _0^{*}+\chi _{c\bar{c}}V_p\Phi _0, & \chi _{\bar{c}%
c}V_p\Phi _0^{*}+\chi _{\bar{c}\bar{c}}V_p\Phi _0
\end{array}
\right] ^T$ and $\left[
\begin{array}{lll}
\chi _{\tilde{n}\tilde{n}}V_p, & \chi _{\tilde{m}\tilde{n}}V_p, & \chi _{%
\tilde{m}^{+}\tilde{n}}V_p
\end{array}
\right] ^T$ on the right-hand side of Eqs. (\ref{HFBRPA1}) and (\ref{HFBRPA2}%
), respectively. The various density fluctuations are then
calculated by the method of Capuzzi and Hern\'{a}ndez
\cite{capuzzi}, with a discretization of the dynamical equations
on a spatial mesh of up to 256 points. The frequencies of the
collective excitations of the system can be extracted from the
resonances of the spectral function $\chi ^{^{\prime \prime
}}\left( \omega \right) $, which is also the quantity of
experimental interest. This is defined in the HFB-RPA as
\begin{equation}
\chi ^{^{\prime \prime }}\left( \omega \right) =\chi _C^{^{\prime
\prime }}\left( \omega \right) +\chi _T^{^{\prime \prime }}\left(
\omega \right) \label{kapatt}
\end{equation}
where
\begin{equation}
\chi _C^{^{\prime \prime }}\left( \omega \right) =-\frac 1\pi Im\int d%
\mathbf{r}V_p(\mathbf{r})\left( \Phi _0^{*}\delta \Phi +\Phi
_0\delta \Phi ^{*}\right)   \label{spcc}
\end{equation}
and
\begin{equation}
\chi _T^{^{\prime \prime }}\left( \omega \right) =-\frac 1\pi Im\int d%
\mathbf{r}V_p(\mathbf{r})\left( \delta \tilde{n}+\delta
\tilde{m}+\delta \tilde{m}^{*}\right) .  \label{sptt}
\end{equation}
Here the indices $C$ and $T$ refer to the contributions from the
condensate and from the non-condensate. Other quantities of
interest are the density fluctuations of the condensate and the
non-condensate, which are readily extracted from the solution of
Eqs. (\ref{HFBRPA1}) and (\ref{HFBRPA2}).

The main technical difficulty in the numerical calculations is how
to renormalize the ultraviolet divergence caused by the use of
contact interactions \cite{huang}. The divergence appears in the
equilibrium anomalous density $\tilde{m}^0\left( \mathbf{r}\right)
$ and in the response
functions $\chi _{\tilde{m}\tilde{m}^{+}}$ and $\chi _{\tilde{m}^{+}\tilde{m}%
}$. The simplest way to implement renormalization is by removing
the zero-temperature component of the above quantities. This
procedure is not fully correct as it neglects the quantum
contributions \cite{hutchinson}, but these are extremely small at
temperatures where the thermal corrections become important.
Alternatively one can apply renormalization by
regularizing $\tilde{m}^0\left( \mathbf{r}\right) $, $\chi _{\tilde{m}\tilde{%
m}^{+}}$ and $\chi _{\tilde{m}^{+}\tilde{m}}$ in real space
\cite{bruun}. We have checked that these two procedures give
almost the same mode frequencies in calculations based on the
standard SOBP theory.

In brief, the numerical method that we have used consists of three
steps. Firstly, we solve the HFB Eqs. (\ref{GPE}) and (\ref{BdG})
(or, in case of HF-RPA, the corresponding HF equations) to
determine the equilibrium densities and quasiparticle amplitudes.
We then construct the bare two-particle response functions and
compute the dynamic fluctuations from Eqs. (\ref{HFBRPA1}) and
(\ref{HFBRPA2}). We finally calculate the imaginary part of the
response functions according to Eq. (\ref{kapatt}). In the present
case of an isotropic trap, the calculations can be greatly
simplified by projecting the RPA equations and the response
functions onto the various multipole modes \cite{capuzzi}. We
shall be interested in the
monopolar, dipolar, and quadrupolar excitations, which require setting $%
V_p\left( \mathbf{r}\right) \propto r^2$,
$V_p\left(\mathbf{r}\right) \propto r\cos \theta $ and $V_p\left(
\mathbf{r}\right) \propto r^2Y_{20}\left( \theta ,\varphi \right)
$.

In the following we evaluate a gas of $N=2000$ $^{87}$Rb atoms in
a spherical trap with trap frequency $\omega _0=2\pi \times 182.5$
Hz, this value being the geometric average of the axial and radial
frequencies in the JILA experiments \cite{jin97}. The temperature
is taken in units of the critical temperature for an ideal gas
with the same value of $N$ and $\omega _0$, which is
$T_c=0.94\hbar \omega _0N^{1/3}$. In most calculations we use a
basis of $n\leq n_{\max }=24$ and $l\leq l_{\max }=32$ for the
quasiparticle wavefunctions, where the indices $n$ and $l$ label
the number of radial nodes and the orbital angular momentum of the
wavefunction.

\begin{figure}[tbp]
\centerline{\includegraphics[width=10.0cm,angle=-90,clip=]{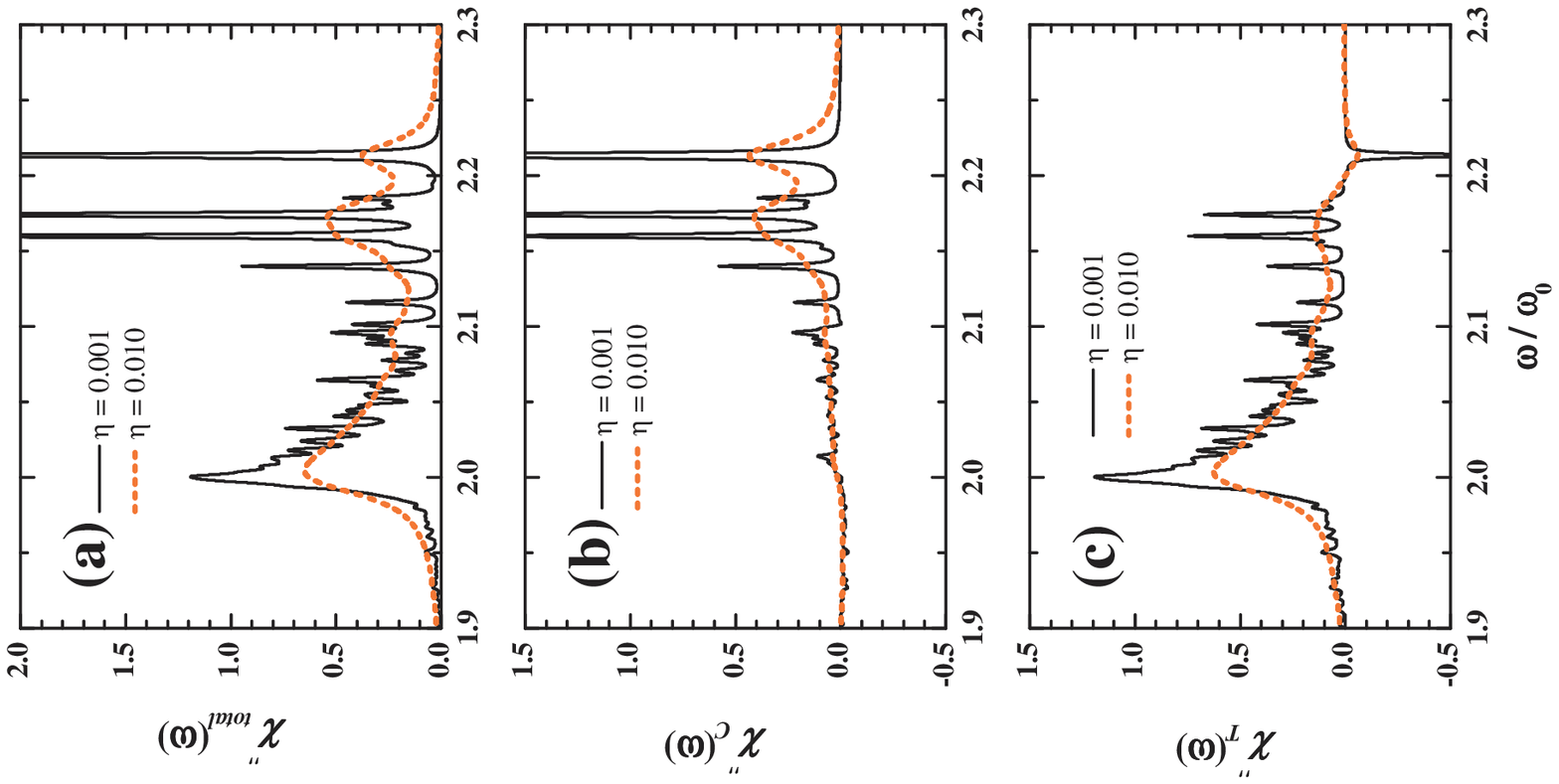}}
\caption{Spectral responses (in arbitrary units) for the monopolar
excitation as functions of frequency $\omega$ (in units of
$\omega_0$) as calculated from the HF-RPA at $T/T_c=0.5$, plotted
for two values of $\eta $ (in units of $\omega_0$) as indicated in
the panels. The three panels display the total spectral response
(a) and the contributions of the condensate (b) and of the
non-condensate (c).} \label{fig1}
\end{figure}

\begin{figure}[tbp]
\centerline{\includegraphics[width=6.0cm,angle=-90,clip=]{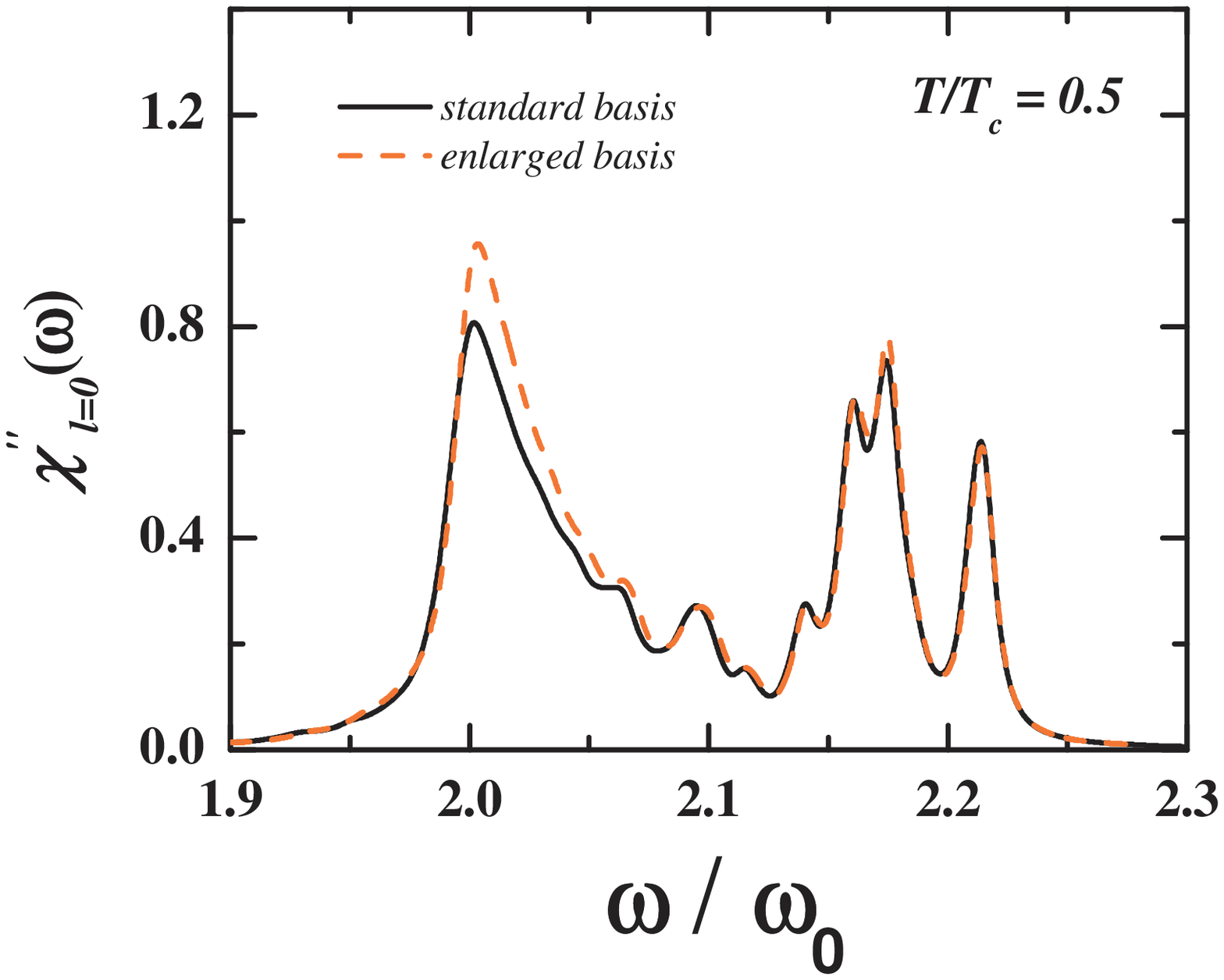}}
\caption{Spectral response (in arbitrary units) for the monopolar
excitation as a function of frequency $\omega$ (in units of
$\omega_0$) as calculated from the HF-RPA at $T/T_c=0.5$ and $\eta
= 0.005 \omega_0$ with two kinds of basis set.} \label{fig2}
\end{figure}

\subsection{Tests of numerical accuracy}

In this subsection we report some tests of the accuracy of our
numerical calculations. First of all, we must replace the positive
infinitesimal quantity $\eta $ in the reference response functions
by a finite value. In Fig. 1 we show the spectral functions for
the monopolar excitation in the HF-RPA for two values of $\eta $
at a reduced temperature $T/T_c=0.5$. For small value of $\eta $
many spikes appear in the spectrum, due to the discrete basis set
that was chosen for the dynamical description. With increasing
$\eta $ these spikes are rounded off into broad resonances, which
are insensitive to the precise value of $\eta $. In the following
we preferentially take $\eta =0.01\omega _0$ in calculating the
spectral functions, this choice being consistent with a typical
experimental energy resolution \cite{jin97}.

The other aspect of the calculations that needs examining is the
role of the basis set. In Fig. 2 we show the HF-RPA monopole
spectrum at $T/T_c=0.6$ and $\eta =0.005\omega _0$, as calculated
from two choices of basis set. These are the standard set as
described above (solid line) and a set in which the number of
basis function has been doubled (dashed line). No quantitative
changes are found for the condensate response around $\omega
=2.2\omega _0$, while for the response of the thermal cloud near
$\omega =2.0\omega _0$ only a small change is present in the
spectral intensity.

\begin{figure}[tbp]
\centerline{\includegraphics[width=10.0cm,angle=-90,clip=]{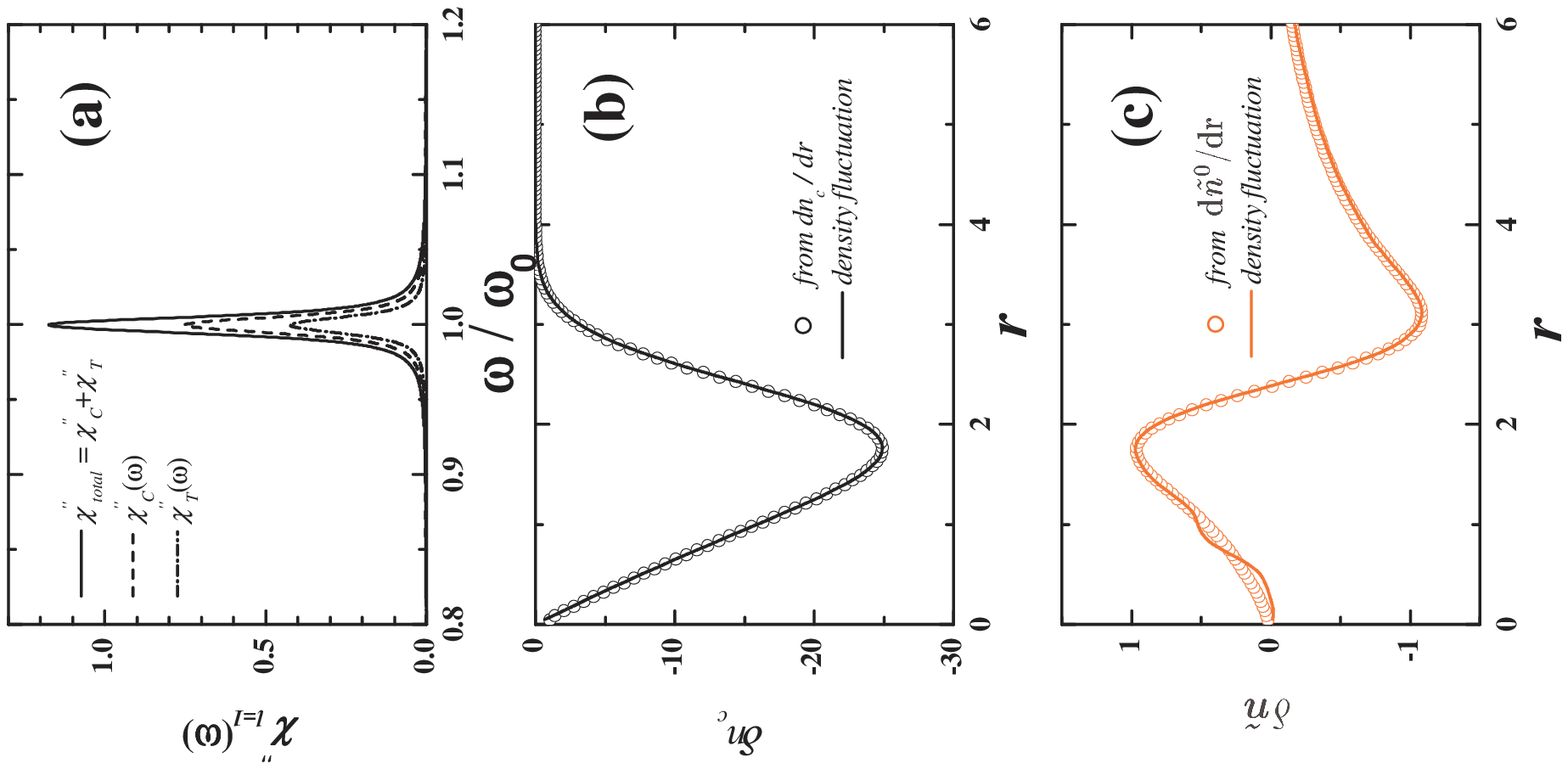}}
\caption{(a) Spectral response (in arbitrary units) for the
dipolar excitation as a function of frequency $\omega$ (in units
of $\omega_0$), as calculated from the HF-RPA at $T/T_c = 0.6 $
with the choice $\eta = 0.005 \omega_0$. The density fluctuations
at resonance (in arbitrary units) are plotted as functions of the
radial coordinate $r$ (in units of $a_{ho}=(\hbar /m\omega
_0)^{1/2}$) in (b) for the condensate and in (c) for the
non-condensate (solid lines). In the same panels are also shown
the corresponding results from the analytical expressions of the
mode eigenvectors (circles).} \label{fig3}
\end{figure}

\begin{figure}[tbp]
\centerline{\includegraphics[width=10.0cm,angle=-90,clip=]{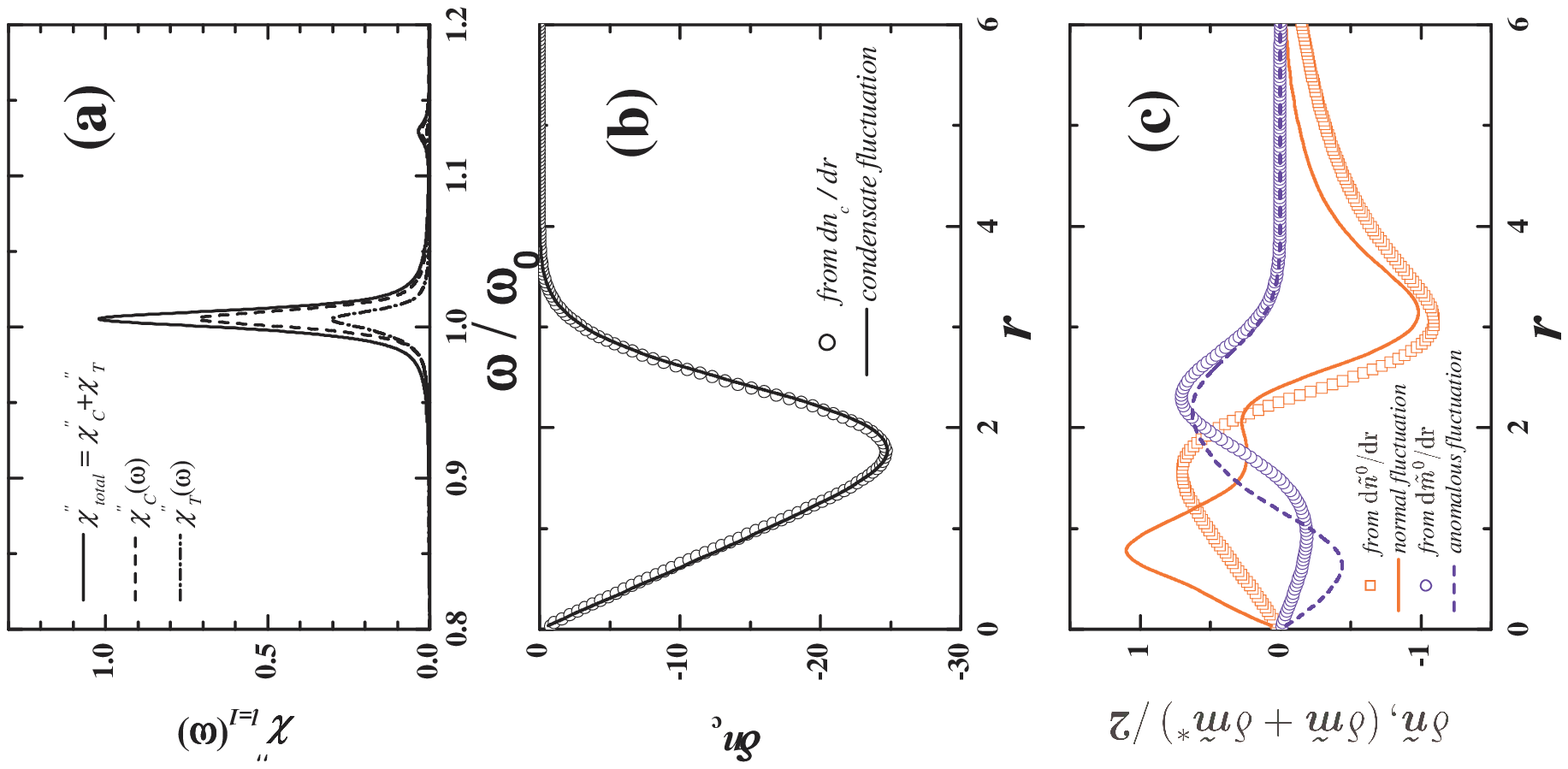}}
\caption{The same as Fig. 3 for the HFB-RPA.} \label{fig4}
\end{figure}

\section{Dipole mode}

An important check on the accuracy of the theory is offered by the
Kohn
theorem. One can analytically prove that the dipolar oscillation in the $%
\alpha $ direction (with $\alpha =x$, $y$, or $z$ in the general
case of an anisotropic trap) is described by the Ansatz $\delta
\Phi =\left( \partial
/\partial r_\alpha -mr_\alpha \omega _\alpha /\hbar \right) \Phi _0$, $%
\delta \Phi ^{*}=\left( \partial /\partial r_\alpha +mr_\alpha
\omega
_\alpha /\hbar \right) \Phi _0^{*}$, $\delta \tilde{n}=\partial \tilde{n}%
^0/\partial r_\alpha $, $\delta \tilde{m}=\left( \partial
/\partial r_\alpha
-2mr_\alpha \omega _\alpha /\hbar \right) \tilde{m}^0$, and $\delta \tilde{m}%
^{*}=\left( \partial /\partial r_\alpha +2mr_\alpha \omega _\alpha
/\hbar \right) \tilde{m}^{0*}$. The theorem asserts that the
corresponding mode frequency is given by the bare trap frequency
$\omega _\alpha $.

In Fig. 3(a) we show the spectral response for a dipolar
excitation as obtained from the HF-RPA at $T/T_c=0.6$ and $\eta
=0.005\omega _0$. It has been explicitly shown that the Kohn
theorem is satisfied in this approach \cite{reidl01,minguzzi}. As
a result a sharp resonance is present in the HF-RPA dipole
spectrum at $\omega =\omega _0$. The density fluctuations at the
resonance, as calculated from the solution of the dynamical
equations, are plotted in Figs. 3(b) and 3(c) as solid lines and
are compared with the predictions of the above Ansatz (circles).
The two methods give almost the same result for both condensate
and thermal density fluctuations, except for a weak structure in
the thermal density fluctuation which may be due to the truncation
of the basis sets.

In Fig. 4 we show the spectral response of the dipole mode as
obtained from the HFB-RPA with the same choice of parameters. In
this approximation, the generalized Kohn theorem is not exactly
satisfied, since a secondary peak is found in the spectrum at
$\omega \simeq 1.13\omega _0$. According to the discussion given
by Lewenstein and You \cite{you}, a possible reason for this
inaccuracy is the non-completeness of the set of quasiparticle
wavefunctions used in the calculation. There also are appreciable
distortions of the eigenvectors for the non-condensate
oscillations in Fig. 4(c).

\section{Monopole and quadrupole modes}

\begin{figure}[tbp]
\centerline{\includegraphics[width=6.0cm,angle=-90,clip=]{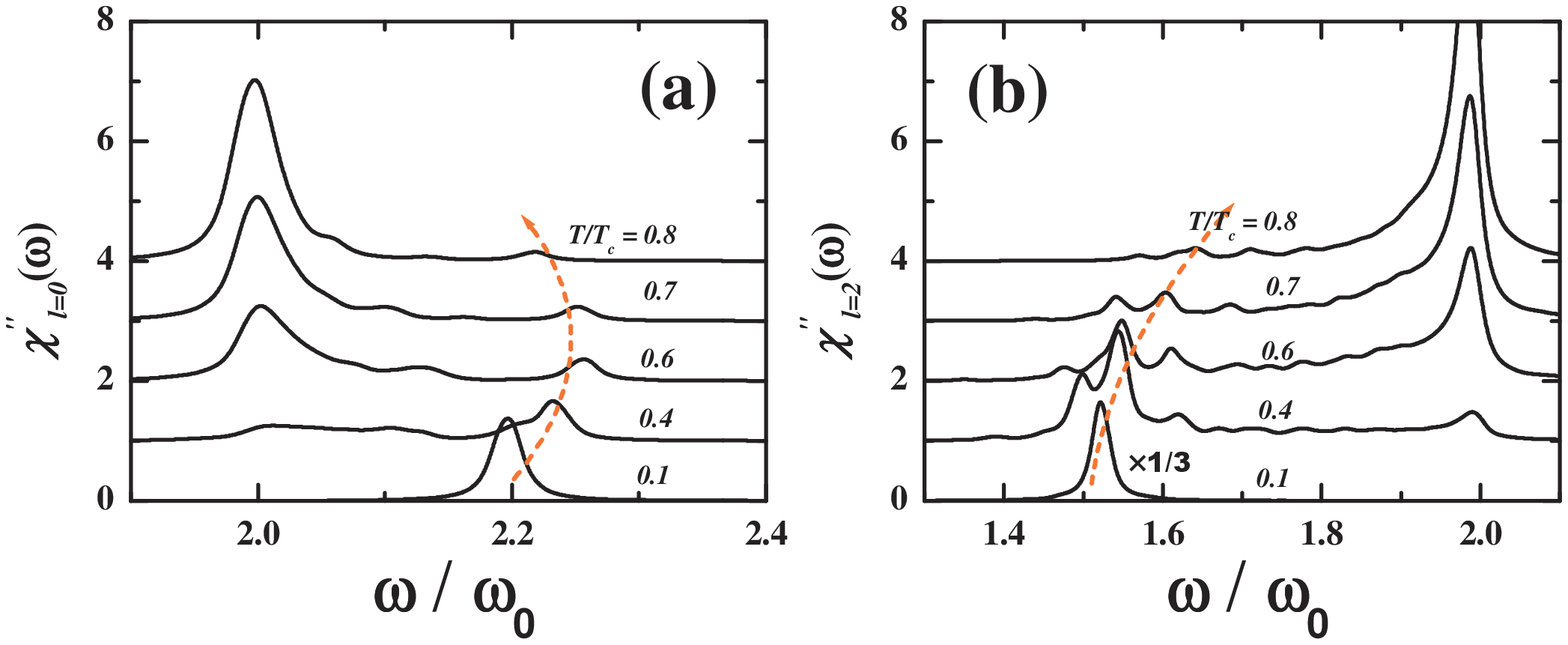}}
\caption{Spectral response (in arbitrary units) as a function of
frequency $\omega$ (in units of $\omega_0$) for the monopole mode
(a)and the quadrupole mode (b), as calculated with $\eta =
0.01\omega_0$ from the HFB-RPA at the temperatures indicated in
the figure. The curves are progressively shifted upwards by one
unit for clarity and the quadrupole response at $T/T_c = 0.1$ is
reduced by a factor of 3. The dashed line in each panel indicates
how the condensate resonance moves with temperature.} \label{fig5}
\end{figure}

\begin{figure}[tbp]
\centerline{\includegraphics[width=6.0cm,angle=-90,clip=]{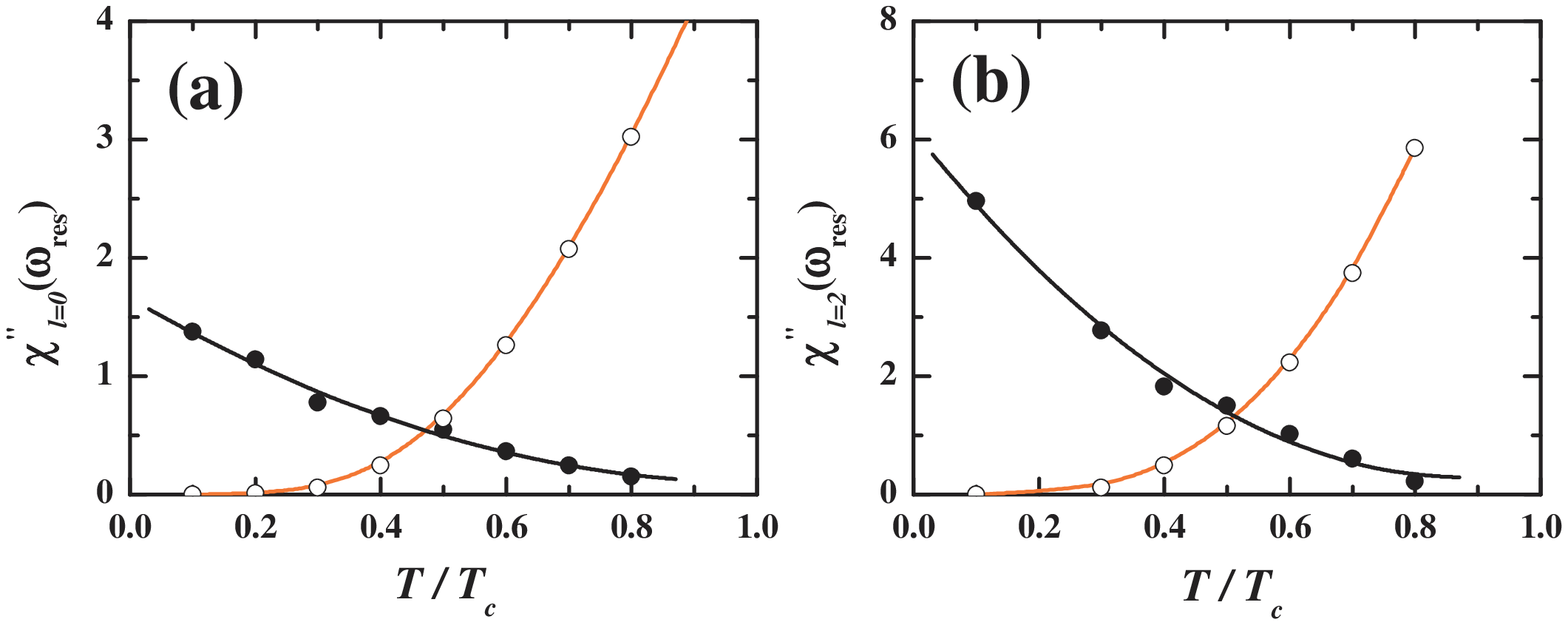}}
\caption{Amplitude of the HFB-RPA resonances (in arbitrary units)
from Fig. 5 as a function of reduced temperature $T/T_c$ for the
monopole (a) and the quadrupole (b). The solid and empty circles
refer to the condensate and to the non-condensate, respectively.
The lines are guides to the eye.} \label{fig6}
\end{figure}

We present in this section the numerical results of the HFB-RPA
for the monopole and quadrupole modes and compare them with those
given by the SOBP-RPA and by the HF-RPA. These various theories
give somewhat different results for the spectra at intermediate
values of the temperature, in the range $0.4T_c\leq T\leq 0.8T_c$.

In Fig. 5 we plot the HFB-RPA spectral functions at various
temperatures. For $k_BT\gtrsim \mu $ two main resonances are seen
in each spectrum, which can be interpreted as representing the
collective oscillations of the non-condensate and of the
condensate. The oscillator strength of each resonance has been
extracted from the spectra and is shown in Fig. 6 as a function of
temperature. Naturally, with increasing $T/T_c$ the amplitude of
the non-condensate resonances grows (empty  circles) while that of
the condensate resonances decreases (solid circles). The
amplitudes of the modes
in the two components of the gas are comparable with each other near $%
T/T_c=0.5$, where the non-condensate fraction is populated by
about $30\%$ for our choice of parameters. Above this temperature
the strength of the non-condensate resonances increases very
rapidly.

\begin{figure}[tbp]
\centerline{\includegraphics[width=6.0cm,angle=-90,clip=]{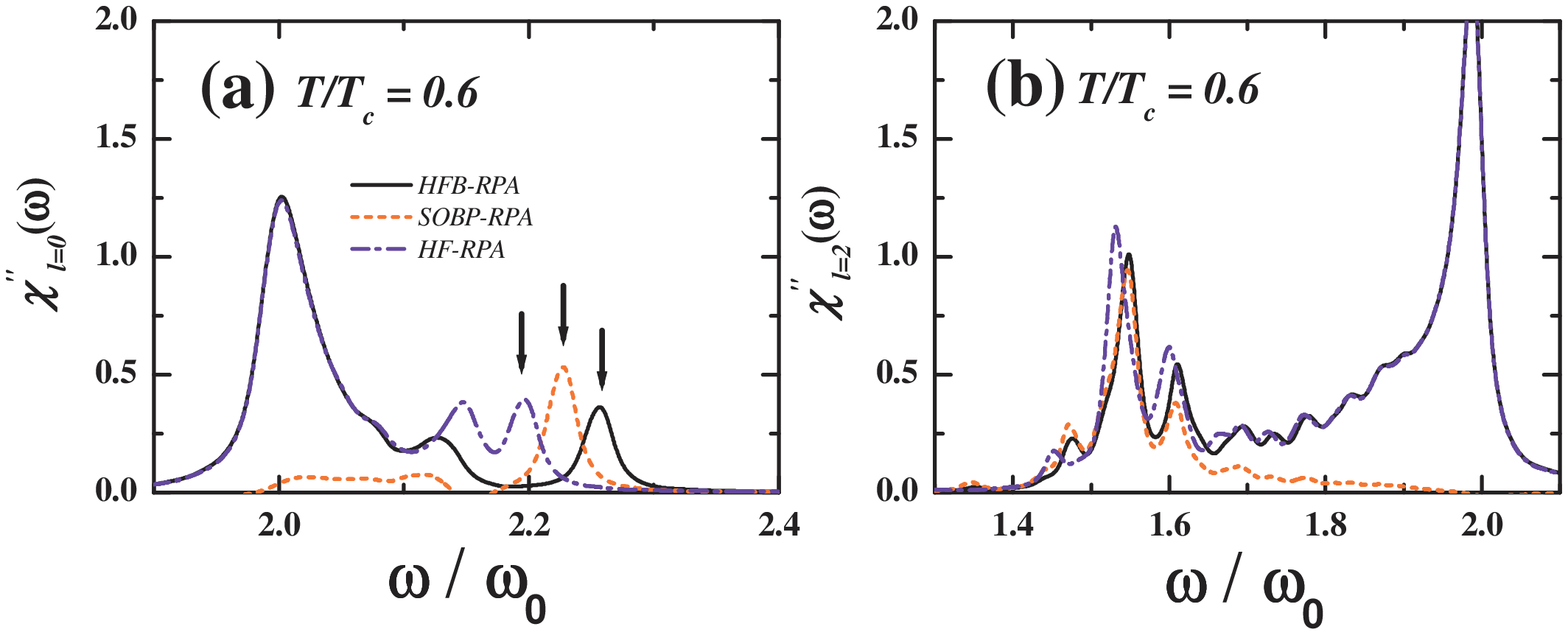}}
\caption{Spectral response (in arbitrary units) as a function of
frequency $\omega$ (in units of $\omega_0$) for the monopole mode
(a) and the quadrupole mode (b), at $T/T_c = 0.6$ with $\eta =
0.01\omega_0$ from the HFB-RPA (solid lines), the SOBP-RPA (dashed
lines) and the HF-RPA (dot-dashed lines). The arrows in panel (a)
point to the condensate resonance position given by each RPA
theory. The SOBP-RPA spectra as defined in Eqs. (\ref{sobp1}) and
(\ref{matrixd}) do not include the contribution from the direct
excitation of the non-condensate. } \label{fig7}
\end{figure}

In Fig. 7 we compare with each other the numerical results from
the RPA theories for the monopolar and quadrupolar spectra at
$T/T_c = 0.6$. We see that the HF-RPA and HFB-RPA closely agree in
their predictions on the main non-condensate resonances for both
types of excitations. We also see that all three theories predict
essentially very similar results for the main quadrupolar
resonance of the condensate, the position of the main peak at
$\omega \simeq 1.55\omega_0$ in Fig. 7(b) being also in agreement
with the result of the HFB-Popov approximation (not shown). In the
following we concentrate on the main condensate resonance in the
monopolar mode, for which the three theories give rather different
predictions as is emphasized by the three arrows in Fig. 7(a). In
fact, the partial spectra of condensate and non-condensate show an
appreciable overlap in this frequency range, implying a stronger
dynamical coupling between the breathing excitations of the two
components of the gas and therefore an enhanced sensitivity to the
approximations made in the theory.

\begin{figure}[tbp]
\centerline{\includegraphics[width=8.0cm,angle=-90,clip=]{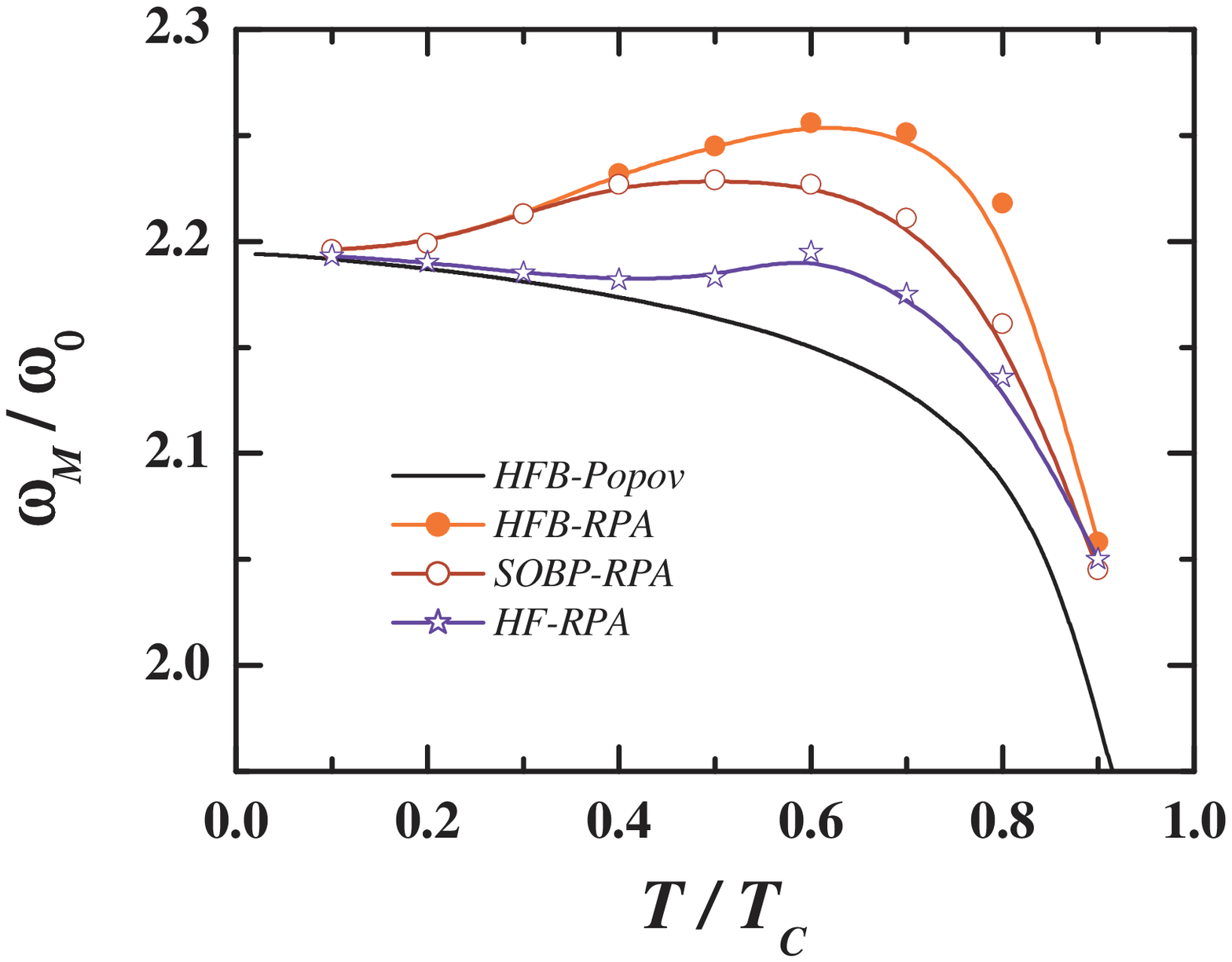}}
\caption{Monopole excitation frequency $\omega_M$ (in units of
$\omega_0$) as a function of reduced temperature $T/T_c$, as
predicted by various theories: the HFB-Popov (solid line), the
HFB-RPA (solid circles), the SOBP-RPA (empty circles) and the
HF-RPA (stars). The lines connecting the symbols are guides to the
eye.} \label{fig8}
\end{figure}

To better illustrate the difference among the various theories, we
extract the monopolar mode frequency of the condensate from the
peak in $\chi ^{^{\prime \prime }}\left( \omega \right) $ and plot
it in Fig. 8 as a function of reduced temperature. For comparison
we also show the mode frequency given by the HFB-Popov theory (see
Sec. II). The most remarkable feature of Fig. 8 is that all three
RPA theories show a \emph{non-monotonic} behavior of the resonance
as a function of temperature, in contrast with the prediction of
the HFB-Popov theory in which the resonance frequency decreases
monotonically with  increasing temperature. This difference is due
to the dynamical coupling between the condensate and the
non-condensate, which is neglected in the mean-field theory and
becomes important as the non-condensate is significantly
populated.

Let us now compare the three RPA theories, which transcend the
mean-field level. At low temperature ($T/T_c < 0.4$) we observe
two different trends: the mode frequencies obtained from the
HFB-RPA and from the SOBP-RPA are in close agreement and move
upwards with temperature, whereas the mode frequency predicted by
the HF-RPA tends to decrease. The latter trend is in good
agreement with the HFB-Popov theory, in accord with the proof
already given in Ref. \cite{mt}. The upward trend of the mode
frequency with temperature is manifested in all RPA theories at
intermediate temperatures, reaching near $T/T_c=0.7$ the highest
sensitivity to the detailed description of the physical process in
which the thermal cloud is driven by its self-generated dynamical
potential. Finally, in proximity of the critical temperature all
three theories tend to agree as the anomalous density fluctuations
disappear.

\begin{figure}[tbp]
\centerline{\includegraphics[width=6.0cm,angle=-90,clip=]{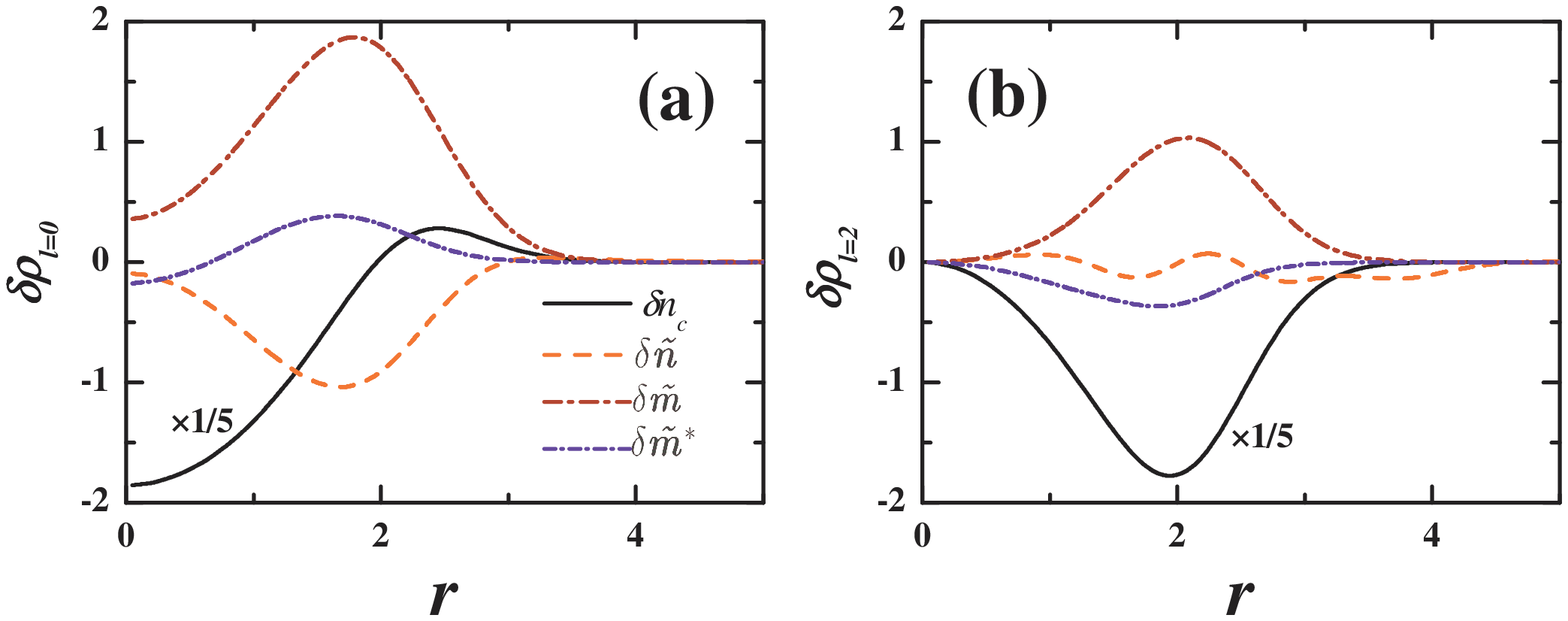}}
\caption{Density fluctuations (in arbitrary units) as functions of
the radial coordinate $r$ (in units of $a_{ho}$) for the monopole
mode (a) and the quadrupole mode (b), as calculated from the
HFB-RPA for $T/T_c = 0.6$ at the appropriate excitation frequency
of the condensate. In both panels the condensate density
fluctuation is reduced by a factor of 5 for clarity.} \label{fig9}
\end{figure}

The fact that a large upward frequency shift is found with
increasing temperature in both the SOBP-RPA and the HFB-RPA
suggests that a significant role is played by the anomalous
density fluctuations. In Fig. 9 we show the partial density
fluctuations which accompany the monopolar and quadrupolar
condensate resonances at $T/T_c=0.6$, as calculated from the
HFB-RPA. In both modes we find that the anomalous density
fluctuations are at this temperature at least comparable in
magnitude to the fluctuations of the normal density.

\section{Conclusions}

In conclusion, we have developed a random-phase theory for the
dynamics of a weakly interacting Bose gas under external
confinement at finite temperature. In the theory the dynamics of
the condensate and of the thermal cloud are treated on the same
footing and a previous Hartree-Fock random-phase scheme is
extended through the inclusion of the anomalous density
fluctuations.  The theory satisfies with good numerical accuracy
the generalized Kohn theorem and correctly reduces to the
second-order Beliaev-Popov theory if one neglects the process in
which the thermal cloud is driven by its self-generated potential.
It thereby fully includes the Landau-Beliaev damping mechanism.

We have compared the theory with the second-order Beliaev-Popov
theory and with the Hartree-Fock random-phase theory by numerical
illustrations for a condensate of $^{87}$Rb atoms inside a
spherical trap. The location of the main monopolar and quadrupolar
resonances of the thermal cloud are well reproduced in the
Hartree-Fock RPA and the frequency of the quadrupole mode of the
condensate does not differ significantly from the mean-field
HFB-Popov prediction. We have instead found that for $T > 0.4T_c$
the temperature dependence of the breathing mode frequency of the
condensate obtained from the various RPA theories is very
different from the HFB-Popov result. A significant role appears to
be played in the dynamics of the Bose-condensed gas by the
anomalous density fluctuations of the thermal cloud at
intermediate temperatures, even though they are known not to
affect significantly the thermodynamics of the trapped gas
\cite{minguzzi97,giorgini97,guilleumas}.

Our results, though restricted to isotropic confinement, may be
relevant in connection with the JILA experiments \cite{jin97},
where the breathing mode in an anisotropic trap showed a frequency
upshift with temperature which could not be accounted for by the
HFB-Popov theory \cite{burnett}. A quantitative comparison between
experimental data and the RPA predictions for an anisotropic trap
would be interesting for a full test of the theory and we hope to
address this issue in future work.

\begin{acknowledgments}
This work has been partially supported by INFM under the
PRA-Photonmatter Programme. One of the authors (X.-J. L.) was
supported by the NSF-China under Grant No. 10205022 and by the
National Fundamental Research Program (NFRP) under Grant No.
001CB309308.
\end{acknowledgments}

\begin{appendix}
\section{The two-particle response functions}

We present here a brief explanation on how to derive the response
functions used in Eqs. (\ref{HFBRPA1}) and (\ref{HFBRPA2}) and
list the two-particle response functions of the non-condensate.

Let us consider for example the expression of $\chi _{cc}\left( \mathbf{r},%
\mathbf{r}^{\prime };\omega \right) $. The most convenient way to
obtain it is to calculate the bosonic Matsubara Green's function
with imaginary time variable \cite{fetter},
\begin{equation}
\chi _{cc}\left( \mathbf{r},\mathbf{r}^{\prime };\tau \right)
=-\left\langle \text{T}_\tau \tilde{\psi}(\mathbf{r},\tau
)\tilde{\psi}(\mathbf{r}^{\prime },0)\right\rangle _0.  \label{A1}
\end{equation}
Here T$_\tau $ denote the ordering in imaginary time and
$\left\langle ...\right\rangle _0$ denotes the equilibrium
statistical average. By expressing the operator
$\tilde{\psi}(\mathbf{r},\tau )$ in terms of the
Bogoliubov quasiparticle operators $\hat{\alpha}_i$ and $\hat{\alpha}_i^{+}$%
, $\tilde{\psi}(\mathbf{r},\tau )=\sum_j\left[ u_j(\mathbf{r})\hat{\alpha}%
_je^{-\epsilon _j\tau
}+v_j^{*}(\mathbf{r})\hat{\alpha}_j^{+}e^{\epsilon
_j\tau }\right] $, we can rewrite $\chi _{cc}\left( \mathbf{r},\mathbf{r}%
^{\prime };\tau \right) $ in the form
\begin{eqnarray}
\chi _{cc}\left( \mathbf{r},\mathbf{r}^{\prime };\tau \geq
0\right)
&=&-\left\langle \tilde{\psi}(\mathbf{r},\tau )\tilde{\psi}^{+}(\mathbf{r}%
^{\prime },0)\right\rangle _0,  \nonumber \\
&=&-\sum_{j,k}\left\langle \left[
u_j(\mathbf{r})\hat{\alpha}_je^{-\epsilon _j\tau
}+v_j^{*}(\mathbf{r})\hat{\alpha}_j^{+}e^{\epsilon _j\tau }\right]
\left[ u_k(\mathbf{r}^{\prime })\hat{\alpha}_k+v_k^{*}(\mathbf{r}^{\prime })%
\hat{\alpha}_k^{+}\right] \right\rangle _0,  \nonumber \\
&=&-\sum_j\left[ u_j(\mathbf{r})v_j^{*}(\mathbf{r}^{\prime
})\left( 1+f_j\right) e^{-\epsilon _j\tau
}+v_j^{*}(\mathbf{r})u_j(\mathbf{r}^{\prime })f_je^{\epsilon
_j\tau }\right] .  \label{A2}
\end{eqnarray}
We then carry out a Fourier transform with respect to the
imaginary time variable $\tau $,
\begin{eqnarray}
\chi _{cc}\left( \mathbf{r},\mathbf{r}^{\prime };i\omega _n\right)
&=&\int\limits_0^\beta d\tau e^{i\omega _n\tau }\chi _{cc}\left( \mathbf{r},%
\mathbf{r}^{\prime };\tau \geq 0\right) ,  \nonumber \\
&=&\sum_j\left( \frac{u_j\left( \mathbf{r}\right) v_j^{*}\left( \mathbf{r}%
^{\prime }\right) }{i\omega _n-\epsilon _j}-\frac{v_j^{*}\left( \mathbf{r}%
\right) u_j\left( \mathbf{r}^{\prime }\right) }{i\omega _n+\epsilon _j}%
\right)  \label{A3}
\end{eqnarray}
where $i\omega _n=2n\pi i/\beta $. With the analytic continuation
$i\omega _n\rightarrow \omega +i\eta $ we obtain the expression
for $\chi _{cc}\left( \mathbf{r},\mathbf{r}^{\prime };\omega
\right) $ in Eq. (\ref{kapacccc}).

The two-particle response functions of the non-condensate can be
derived in a similar way. They take the following forms:
\begin{equation}
\chi _{\tilde{n}\tilde{n}}\left( \mathbf{r},\mathbf{r}^{\prime
};\omega
\right) =\chi _{\tilde{n}\tilde{n}}^{(1)}\left( \mathbf{r},\mathbf{r}%
^{\prime };\omega \right) +\chi _{\tilde{n}\tilde{n}}^{(2)}\left( \mathbf{r},%
\mathbf{r}^{\prime };\omega \right)   \label{A4}
\end{equation}
with
\[
\chi _{\tilde{n}\tilde{n}}^{(1)}\left(
\mathbf{r},\mathbf{r}^{\prime };\omega \right)
=\sum\limits_{ij}\frac{\left( u_i^{*}u_j+v_i^{*}v_j\right) \left(
u_iu_j^{*}+v_iv_j^{*}\right) \left( f_i-f_j\right) }{\hbar \omega
^{+}+\left( \epsilon _i-\epsilon _j\right) }
\]
and
\begin{eqnarray*}
\chi _{\tilde{n}\tilde{n}}^{(2)}\left(
\mathbf{r},\mathbf{r}^{\prime };\omega \right)  &=&\frac
12\sum\limits_{ij}\left[ \frac{\left( u_iv_j+v_iu_j\right) \left(
u_i^{*}v_j^{*}+v_i^{*}u_j^{*}\right) \left(
1+f_i+f_j\right) }{\hbar \omega ^{+}-\left( \epsilon _i+\epsilon _j\right) }%
\right.  \\
&&\left. -\frac{\left( u_i^{*}v_j^{*}+v_i^{*}u_j^{*}\right) \left(
u_iv_j+v_iu_j\right) \left( 1+f_i+f_j\right) }{\hbar \omega
^{+}+\left( \epsilon _i+\epsilon _j\right) }\right] ;
\end{eqnarray*}
\begin{equation}
\chi _{\tilde{n}\tilde{m}}\left( \mathbf{r},\mathbf{r}^{\prime
};\omega
\right) =\chi _{\tilde{m}^{+}\tilde{n}}^{*}\left( \mathbf{r}^{\prime },%
\mathbf{r};\omega \right) =\chi _{\tilde{n}\tilde{m}}^{(1)}\left( \mathbf{r},%
\mathbf{r}^{\prime };\omega \right) +\chi
_{\tilde{n}\tilde{m}}^{(2)}\left( \mathbf{r},\mathbf{r}^{\prime
};\omega \right)   \label{A5}
\end{equation}
with
\[
\chi _{\tilde{n}\tilde{m}}^{(1)}\left(
\mathbf{r},\mathbf{r}^{\prime };\omega \right)
=2\sum\limits_{ij}\frac{\left( u_i^{*}u_j+v_i^{*}v_j\right)
u_iv_j^{*}\left( f_i-f_j\right) }{\hbar \omega ^{+}+\left(
\epsilon _i-\epsilon _j\right) }
\]
and
\[
\chi _{\tilde{n}\tilde{m}}^{(2)}\left(
\mathbf{r},\mathbf{r}^{\prime };\omega \right)
=2\sum\limits_{ij}\left[ \frac{v_iu_jv_i^{*}v_j^{*}\left(
1+f_i+f_j\right) }{\hbar \omega ^{+}-\left( \epsilon _i+\epsilon _j\right) }-%
\frac{u_i^{*}v_j^{*}u_iu_j\left( 1+f_i+f_j\right) }{\hbar \omega
^{+}+\left( \epsilon _i+\epsilon _j\right) }\right] ;
\]
\begin{equation}
\chi _{\tilde{n}\tilde{m}^{+}}\left( \mathbf{r},\mathbf{r}^{\prime
};\omega
\right) =\chi _{\tilde{m}\tilde{n}}^{*}\left( \mathbf{r}^{\prime },\mathbf{r}%
;\omega \right) =\chi _{\tilde{n}\tilde{m}^{+}}^{(1)}\left( \mathbf{r},%
\mathbf{r}^{\prime };\omega \right) +\chi _{\tilde{n}\tilde{m}%
^{+}}^{(2)}\left( \mathbf{r},\mathbf{r}^{\prime };\omega \right)
\label{A6}
\end{equation}
with
\[
\chi _{\tilde{n}\tilde{m}^{+}}^{(1)}\left(
\mathbf{r},\mathbf{r}^{\prime };\omega \right)
=2\sum\limits_{ij}\frac{\left( u_i^{*}u_j+v_i^{*}v_j\right)
v_iu_j^{*}\left( f_i-f_j\right) }{\hbar \omega ^{+}+\left(
\epsilon _i-\epsilon _j\right) }
\]
and
\[
\chi _{\tilde{n}\tilde{m}^{+}}^{(2)}\left(
\mathbf{r},\mathbf{r}^{\prime };\omega \right)
=2\sum\limits_{ij}\left[ \frac{v_iu_ju_i^{*}u_j^{*}\left(
1+f_i+f_j\right) }{\hbar \omega ^{+}-\left( \epsilon _i+\epsilon _j\right) }-%
\frac{u_i^{*}v_j^{*}v_iv_j\left( 1+f_i+f_j\right) }{\hbar \omega
^{+}+\left( \epsilon _i+\epsilon _j\right) }\right] ;
\]
\begin{equation}
\chi _{\tilde{m}\tilde{m}}\left( \mathbf{r},\mathbf{r}^{\prime
};\omega
\right) =\chi _{\tilde{m}^{+}\tilde{m}^{+}}^{*}\left( \mathbf{r}^{\prime },%
\mathbf{r};\omega \right) =\chi _{\tilde{m}\tilde{m}}^{(1)}\left( \mathbf{r},%
\mathbf{r}^{\prime };\omega \right) +\chi
_{\tilde{m}\tilde{m}}^{(2)}\left( \mathbf{r},\mathbf{r}^{\prime
};\omega \right)   \label{A7}
\end{equation}
with
\[
\chi _{\tilde{m}\tilde{m}}^{(1)}\left(
\mathbf{r},\mathbf{r}^{\prime };\omega \right)
=4\sum\limits_{ij}\frac{v_i^{*}u_ju_iv_j^{*}\left( f_i-f_j\right)
}{\hbar \omega ^{+}+\left( \epsilon _i-\epsilon _j\right) }
\]
and
\[
\chi _{\tilde{m}\tilde{m}}^{(2)}\left(
\mathbf{r},\mathbf{r}^{\prime };\omega \right)
=2\sum\limits_{ij}\left[ \frac{u_iu_jv_i^{*}v_j^{*}\left(
1+f_i+f_j\right) }{\hbar \omega ^{+}-\left( \epsilon _i+\epsilon _j\right) }-%
\frac{v_i^{*}v_j^{*}u_iu_j\left( 1+f_i+f_j\right) }{\hbar \omega
^{+}+\left( \epsilon _i+\epsilon _j\right) }\right] ;
\]
\begin{equation}
\chi _{\tilde{m}\tilde{m}^{+}}\left( \mathbf{r},\mathbf{r}^{\prime
};\omega
\right) =\chi _{\tilde{m}\tilde{m}^{+}}^{(1)}\left( \mathbf{r},\mathbf{r}%
^{\prime };\omega \right) +\chi
_{\tilde{m}\tilde{m}^{+}}^{(2)}\left(
\mathbf{r},\mathbf{r}^{\prime };\omega \right)   \label{A8}
\end{equation}
with
\[
\chi _{\tilde{m}\tilde{m}^{+}}^{(1)}\left(
\mathbf{r},\mathbf{r}^{\prime };\omega \right)
=4\sum\limits_{ij}\frac{v_i^{*}u_jv_iu_j^{*}\left( f_i-f_j\right)
}{\hbar \omega ^{+}+\left( \epsilon _i-\epsilon _j\right) }
\]
and
\[
\chi _{\tilde{m}\tilde{m}^{+}}^{(2)}\left(
\mathbf{r},\mathbf{r}^{\prime };\omega \right)
=2\sum\limits_{ij}\left[ \frac{u_iu_ju_i^{*}u_j^{*}\left(
1+f_i+f_j\right) }{\hbar \omega ^{+}-\left( \epsilon _i+\epsilon _j\right) }-%
\frac{v_i^{*}v_j^{*}v_iv_j\left( 1+f_i+f_j\right) }{\hbar \omega
^{+}+\left( \epsilon _i+\epsilon _j\right) }\right] ;
\]
and finally
\begin{equation}
\chi _{\tilde{m}^{+}\tilde{m}}\left( \mathbf{r},\mathbf{r}^{\prime
};\omega
\right) =\chi _{\tilde{m}^{+}\tilde{m}}^{(1)}\left( \mathbf{r},\mathbf{r}%
^{\prime };\omega \right) +\chi
_{\tilde{m}^{+}\tilde{m}}^{(2)}\left(
\mathbf{r},\mathbf{r}^{\prime };\omega \right)   \label{A9}
\end{equation}
with
\[
\chi _{\tilde{m}^{+}\tilde{m}}^{(1)}\left(
\mathbf{r},\mathbf{r}^{\prime };\omega \right)
=4\sum\limits_{ij}\frac{u_i^{*}v_ju_iv_j^{*}\left( f_i-f_j\right)
}{\hbar \omega ^{+}+\left( \epsilon _i-\epsilon _j\right) }
\]
and
\[
\chi _{\tilde{m}^{+}\tilde{m}}^{(2)}\left(
\mathbf{r},\mathbf{r}^{\prime };\omega \right)
=2\sum\limits_{ij}\left[ \frac{v_iv_jv_i^{*}v_j^{*}\left(
1+f_i+f_j\right) }{\hbar \omega ^{+}-\left( \epsilon _i+\epsilon _j\right) }-%
\frac{u_i^{*}u_j^{*}u_iu_j\left( 1+f_i+f_j\right) }{\hbar \omega
^{+}+\left( \epsilon _i+\epsilon _j\right) }\right] .
\]
In above expressions $\omega ^{+}=\omega +i0^{+}$ and we have used
abbreviations such as $u_i^{*}u_ju_iu_j^{*}=u_i^{*}(\mathbf{r})u_j(\mathbf{r}%
)u_i(\mathbf{r}^{\prime })u_j^{*}(\mathbf{r}^{\prime })$, which
means that
in the product of four position-dependent functions the first two depend on $%
\mathbf{r}$ and the latter two on $\mathbf{r}^{\prime }$. $\chi
_{ab}^{(1)}$ and $\chi _{ab}^{(2)}$ in the above expressions
correspond to the excitation of single thermal quasiparticles and
of pairs of thermal quasiparticles, respectively.

\end{appendix}


\begin{thebibliography}{99}

\bibitem{jin96}  D. S. Jin, J. R. Ensher, M. R. Matthews, C. E. Wieman, and
E. A. Cornell, Phys. Rev. Lett. {\bf 77}, 420 (1996).

\bibitem{jin97}  D. S. Jin, M. R. Matthews, J. R. Ensher, C. E. Wieman, and
E. A. Cornell, Phys. Rev. Lett. {\bf 78}, 764 (1997).

\bibitem{ketterle96}  M.-O. Mewes, M. R. Andrews, N. J. van Druten, D. M.
Stamper-Kurn, D. S. Durfee, C. G. Townsend, and W. Ketterle, Phys. Rev.
Lett. {\bf 77}, 988 (1996).

\bibitem{ketterle98}  D. M. Stamper-Kurn, H.-J. Miesner, S. Inouye, M. R.
Andrews, and W. Ketterle, Phys. Rev. Lett. {\bf 81}, 500 (1998).

\bibitem{ketterle00}  R. Onofrio, D. S. Durfee, C. Raman, M. K\"{o}hl, C. E. Kuklewicz,
and W. Ketterle, Phys. Rev. Lett. {\bf 84}, 810 (2000).

\bibitem{burnett}  R. J. Dodd, M. Edwards, C. W. Clark, and K. Burnett, Phys. Rev.
A {\bf 57}, R32 (1998).

\bibitem{hutchinson}  D. A. W. Hutchinson, R. J. Dodd, and K. Burnett, Phys.
Rev. Lett. {\bf 81}, 2198 (1998).

\bibitem{overview} For an overview see, A. Griffin
in {\it Bose-Einstein Condensation in Atomic Gases}, Proc. Int.
School of Physics "Enrico Fermi", edited by M. Inguscio, S.
Stringari, and C. E. Wieman (Italian Physical Society, 1999).

\bibitem{mt}  A. Minguzzi and M. P. Tosi, J. Phys.: Condens. Matter {\bf 9},
10211 (1997).

\bibitem{fs}  P. O. Fedichev and G. V. Shlyapnikov, Phys. Rev. A {\bf 58},
3146 (1998).

\bibitem{stoof}  M. J. Bijlsma and H. T. C. Stoof, Phys. Rev. A {\bf 60},
3973 (1999).

\bibitem{zgn}  E. Zaremba, T. Nikuni, and A. Griffin, J. Low Temp. Phys.
{\bf 116}, 277 (1999).

\bibitem{reidl00}  J. Reidl, A. Csord\'{a}s, R. Graham, and P.
Sz\'{e}pfalusy, Phys. Rev. A {\bf 61}, 043606 (2000).

\bibitem{giorgini}  S. Giorgini, Phys. Rev. A {\bf 61}, 063615 (2000).

\bibitem{morgan00}  M. Rusch, S. A. Morgan, D. A. W. Hutchinson, and K.
Burnett, Phys. Rev. Lett. {\bf 85}, 4844 (2000).

\bibitem{morgan03}  S. A. Morgan, M. Rusch, D. A. W. Hutchinson, and K.
Burnett, cond-mat/0305535.

\bibitem{beliaev}  S. T. Beliaev, Sov. Phys. JETP {\bf 7}, 289 (1958); {\bf 7%
}, 299 (1958).

\bibitem{griffin98}  H. Shi and A. Griffin, Phys. Rep. {\bf 304}, 1 (1998).

\bibitem{griffin96}  A. Griffin, Phys. Rev. B {\bf 53}, 9341 (1996).

\bibitem{rmp} F. Dalfovo, S. Giorgini, L. P. Pitaevskii, and S. Stringari,
Rev. Mod. Phys. {\bf 71}, 463 (1999).

\bibitem{liu}  X.-J. Liu and H. Hu, Phys. Rev. A {\bf 68}, 033613
(2003); see also X.-J. Liu {\it et al.} (unpublished).

\bibitem{fetter} A. L. Fetter and J. D. Walecka, {\it Quantum Theory
of Many-Particle Systems} (McGraw-Hill, New York, 1971).

\bibitem{capuzzi}  P. Capuzzi and E. S. Hern\'{a}ndez, Phys. Rev. A {\bf 64}%
, 043607 (2001).

\bibitem{huang} K. Huang, {\it Statistical Mechanics} (Wiley , New York, 1987),
pp. 230-238.

\bibitem{bruun} G. Bruun, Y. Castin, R. Dum, and K. Burnett,
Eur. Phys. J. D {\bf 7}, 433 (1999). The divergence is removed by
the substitutions $\tilde{m}^0\left( {\bf r}\right) \rightarrow
\tilde{m}^0\left( {\bf r}\right) - g\Phi _0^2\left( {\bf r}\right) G_0^{irr}({\bf r%
}) / 2$, $\chi _{\tilde{m}\tilde{m}^{+}}\left( {\bf r},{\bf r}^{\prime %
};\omega \right) \rightarrow \chi _{\tilde{m}\tilde{m}^{+}}\left( {\bf r},{\bf r}%
^{\prime };\omega \right) -\delta ({\bf r}-{\bf r}^{\prime })G_0^{irr}({\bf r%
})$, and $\chi _{\tilde{m}^{+}\tilde{m}}\left( {\bf r},{\bf
r}^{\prime
};\omega \right) \rightarrow \chi _{\tilde{m}^{+}\tilde{m}}\left( {\bf r},%
{\bf r}^{\prime };\omega \right) -\delta ({\bf r}-{\bf r}^{\prime
})G_0^{irr}({\bf r})$. Here $G_0^{irr}({\bf r})$ is the singular
part of the ideal-gas single-particle Green's function $G_0({\bf
r+x}/2,{\bf r-x}/2;\omega =0)$, that diverges as $1/x$ for $x$
$\rightarrow 0$.

\bibitem{reidl01}  J. Reidl, G. Bene, R. Graham, and P. Sz\'{e}pfalusy,
Phys. Rev. A {\bf 63}, 043605 (2001).

\bibitem{minguzzi} A. Minguzzi, Phys. Rev. A {\bf 64}, 033604 (2001).

\bibitem{you} M. Lewenstein and L. You, Phys. Rev. Lett. {\bf 77}, 3489
(1996).

\bibitem{minguzzi97} A. Minguzzi, S. Conti, and M. P. Tosi, J. Phys.: Condens. Matter {\bf 9},
L33 (1997).

\bibitem{giorgini97} S. Giorgini, L. P. Pitaevskii, and S.
Stringari, Phys. Rev. Lett. {\bf 78}, 3987 (1997).

\bibitem{guilleumas} F. Dalfovo, S. Giorgini, M. Guilleumas, L. P. Pitaevskii, and S.
Stringari, Phys. Rev. A {\bf 56}, 3840 (1997).

\end{thebibliography}
\end{document}